\documentclass[aps,prl,superscriptaddress,twocolumn,10pt]{revtex4-2}

\usepackage{siunitx}
\usepackage{physics}
\usepackage{graphicx}
\usepackage{acronym}
\usepackage{xspace}
\usepackage{tikz}
\usepackage{bbold}
\newcommand{\tief}[1]{_\mathrm{#1}}
\newacro{gss}[GS splitting]{ground state splitting}
\newacro{ess}[ES splitting]{excited state splitting}
\newacro{afm}[AFM]{atomic force microscope}
\newacro{siv}[SiV$^-$]{negatively charged silicon vacancy center in diamond}
\newacro{nd}[ND]{nanodiamond}
\newacro{fem}[FEM]{finite element method}
\newcommand{\siv}{\ac{siv}\xspace}
\newcommand{\gss}{\ac{gss}\xspace}
\newcommand{\ess}{\ac{ess}\xspace}
\newcommand{\nd}{\ac{nd}\xspace}
\newcommand{\nds}{\acp{nd}\xspace}
\newcommand{\Timeone}{$T_\mathrm{1}$\xspace}
\newcommand{\Timetwo}{$T_\mathrm{2}$\xspace}
\newcommand{\Timetwostar}{$T_\mathrm{2}^{*}$\xspace}
\newcommand{\Gammaone}{$\Gamma_\mathrm{1}$\xspace}
\begin{document}
	
	
	\title{Prolonged orbital relaxation by locally modified phonon density of states for SiV$^-$ center in nanodiamonds}
	
	
	
	\author{M. Klotz}
	\altaffiliation{These authors contributed equally.}
	\affiliation{Institute for Quantum Optics, Ulm University, 89081 Ulm, Germany}
	\author{K.G. Fehler}
	\altaffiliation{These authors contributed equally.}
	\affiliation{Institute for Quantum Optics, Ulm University, 89081 Ulm, Germany}
	\author{E. S. Steiger}
	\affiliation{Institute for Quantum Optics, Ulm University, 89081 Ulm, Germany}
	\author{S. Häußler}
	\affiliation{Institute for Quantum Optics, Ulm University, 89081 Ulm, Germany}
	\author{R. Waltrich}
	\affiliation{Institute for Quantum Optics, Ulm University, 89081 Ulm, Germany}
	\author{P. Reddy}
	\affiliation{Laser Physics Centre, Research School of Physics, Australian National University, Australian Capital Territory 2601, Australia}
	\author{L. F. Kulikova}
	\affiliation{L.F. Vereshchagin Institute for High Pressure Physics, Russian Academy of Sciences, Troitsk, Moscow 142190, Russia}
	\author{V. A. Davydov}
	\affiliation{L.F. Vereshchagin Institute for High Pressure Physics, Russian Academy of Sciences, Troitsk, Moscow 142190, Russia}
	\author{V. N. Agafonov}
	\affiliation{GREMAN, UMR 7347 CNRS, INSA-CVL, Tours University, 37200 TOURS, France}
	\author{M. W. Doherty}
	\affiliation{Laser Physics Centre, Research School of Physics, Australian National University, Australian Capital Territory 2601, Australia}
	\author{A. Kubanek}
	\affiliation{Institute for Quantum Optics, Ulm University, 89081 Ulm, Germany}
	
	
	\date{\today}
	
	\begin{abstract}
		Coherent quantum systems are a key resource for emerging quantum technology. Solid-state spin systems are of particular importance for compact and scalable devices. However, interaction with the solid-state host degrades the coherence properties. The negatively-charged silicon vacancy center in diamond is such an example. While spectral properties are outstanding, with optical coherence protected by the defects symmetry, the spin coherence is susceptible to rapid orbital relaxation limiting the spin dephasing time. A prolongation of the orbital relaxation time is therefore of utmost urgency and has been tackled by operating at very low temperatures or by introducing large strain. However, both methods have significant drawbacks, the former requires use of dilution refrigerators and the latter affects intrinsic symmetries. Here, a novel method is presented to prolong the orbital relaxation with a locally modified phonon density of states in the relevant frequency range, by restricting the diamond host to below 100 nm. The method works at liquid Helium temperatures of few Kelvin and in the low-strain regime.
	\end{abstract}

	
	\maketitle
	
	In recent years, solid-state quantum emitters have successfully been utilized for applications in quantum information science and sensing. Color centers in diamond, and in particular the \siv, turned out to be a promising system to realize applications that rely on an efficient spin-photon interface and long-lived memories \cite{sukachevSiliconVacancySpinQubit2017}, for example in the context of quantum communication \cite{bhaskarExperimentalDemonstrationMemoryenhanced2020}. However, such sophisticated experiments require to suppress the strong influence from the host environment on the optical and spin coherence. Modifying the interaction with the environment enables to tailor the quantum properties towards the aforementioned applications.
	Here, the group-IV defect centers stand out by their inversion symmetric $D_{3d}$ defect-structures which lead to protection of their optical transitions against charge-fluctuations from e.g. close-by surfaces, when integrated into nano-structures \cite{zhangStronglyCavityEnhancedSpontaneous2018}. As a result, \siv exhibit narrow inhomogeneous line distribution and excellent spectral stability \cite{langLongOpticalCoherence2020}. Due to its relatively small orbital extension as compared to other group IV defects, \siv is highly susceptible to strain introduced by electron-phonon interactions with the environment \cite{meesalaStrainEngineeringSiliconvacancy2018}. On the one hand, this opens up the possibility to use the \siv as a coherent spin-phonon interface for potential transduction between different quantum systems \cite{maityCoherentAcousticControl2020}. On the other hand, the spin coherence time is limited due to fast dephasing processes related to orbital transitions in the groundstate \cite{pingaultCoherentControlSiliconvacancy2017a}. For temperatures below \SI{25}{\kelvin}, the orbital relaxation rate is dominated by a single acoustic phonon process, resonant to the \ac{gss} of $\Delta_\mathrm{GS}/2\pi \approx$ \SI{46}{GHz} as a result of spin-orbit (SO) interaction \cite{jahnkeElectronPhononProcesses2015}.
	Therefore, in high-purity, low-strain bulk diamond the orbital relaxation time (\Timeone) is limited to between \SI{10}{ns} and \SI{40}{ns} for temperatures from \SI{25}{K} to \SI{7}{K}. Hence, decreasing the ground state orbital relaxation rate (\Gammaone) is crucial for future applications that rely on a long spin coherence time (\Timetwo). As a result, one ambition is to increase \Timeone = $\Gamma_\mathrm{1}^{-1}$ so that the spin dephasing time \Timetwostar is no longer limited by orbital relaxation but rather by intrinsic, material related noise sources, such as dipolar coupling to the nuclear spin bath or g-factor fluctuations.
	
	\begin{figure}[t]
		\centering
		\includegraphics[scale=1]{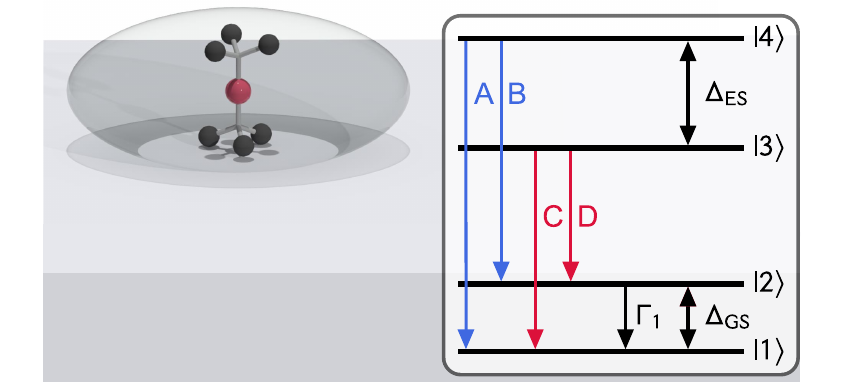}
		\caption{Schematic of a geometrically confined \siv within a \nd placed on a diamond substrate. The defects level structure is also depicted with its zero-phonon-line transitions A, B, C and D. The \gss $\Delta_\mathrm{GS}$ is a result from spin-orbit as well as mechanical interaction with the environment, which leads to an orbital relaxation from $\ket{2} \to \ket{1}$ with rate $\Gamma_1$, where e.g. $\ket{2} = \ket{e_{g_-}\uparrow}$ and $\ket{1} = \ket{e_{g_+}\uparrow}$ in a strain-free environment.}
		\label{fig:sketch}
	\end{figure}
	In general, there are two possibilities to potentially influence \Gammaone. 
	First, decreasing the temperature reduces thermal occupation of phonon modes at the relevant frequencies $\Delta_\mathrm{GS}$ \cite{sukachevSiliconVacancySpinQubit2017}. Experiments have been performed at temperatures of \SI{100}{\milli\kelvin} yielding \Timetwo$ = \SI{13}{\micro\second}$, where $k_\mathrm{B} T \ll \hbar \Delta_\mathrm{GS}$. Already at 2-\SI{4}{K} the spin dephasing time of about \Timetwostar$=\SI{100}{\ns}$ is significantly shortened \cite{sukachevSiliconVacancySpinQubit2017, metschInitializationReadoutNuclear2019}. 
	Second, changing the geometry or applying an external load alters the \siv strain environment and hence locally modifies the spectral coupling density. Imposing an external force has been done by means of a nano-electro-mechanical system, which resulted in prolonged \Timetwostar of \SI{0.25}{\mu s} for a \ac{gss} of \SI{467}{GHz} \cite{sohnControllingCoherenceDiamond2018}. 
	
	
	This work presents an orbital lifetime extension of \siv, that are incorporated into a geometrically confined host, namely a nanodiamond (ND), as illustrated in Fig.\,\ref{fig:sketch}. In order to achieve cold temperatures at the position of the \siv the \nd is in direct thermal contact with a cooled base substrate which inevitably leads to a phononic coupling between the \nd and the base substrate. Numerical simulations of a \nd coupled to a diamond base substrate give detailed insights into the \siv strain response as well as phononic coupling to the substrate. The results are compared with measurements of 14 \nds with varying \gss and temperatures.

	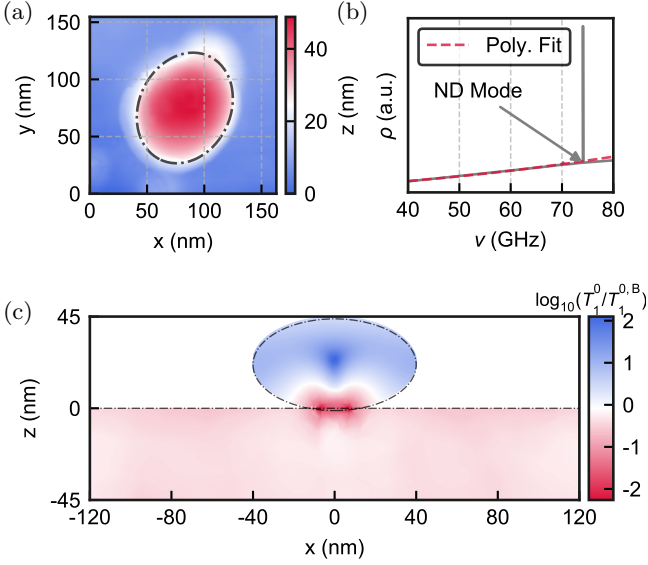
\begin{figure}[t]
		\begin{tikzpicture}
		
		\draw (0, 0) node[inner sep=0] {\includegraphics[scale=.99]{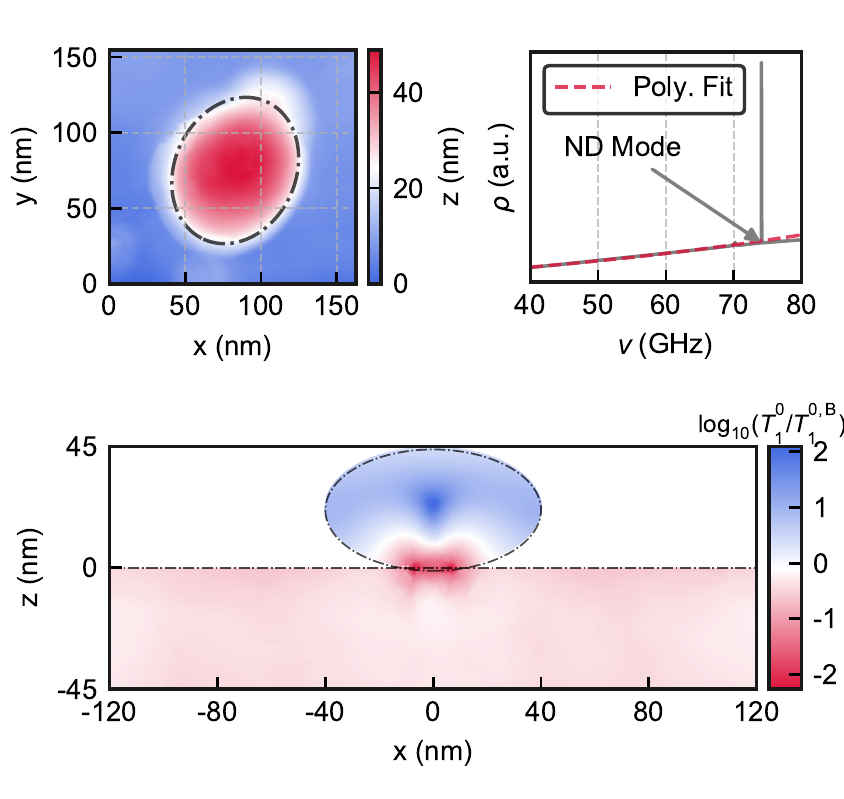}};
		\draw (-116.5pt, 101.5pt) node {(a)};
		\draw (10pt, 101.5pt) node {(b)};
		\draw (-116.5pt, -12pt) node {(c)};
		\end{tikzpicture}
		\caption{(a) AFM scan of a \nd revealing characteristic sizes of $(r_x, r_y, r_z)= (40,50,22.5)$nm, setting an upper size limit for all consecutively measured \nd samples. (b) Density of states (DOS) $\rho$ as a function of frequency $\nu$ for the \nd coupled to bulk with a penetration depth of $d=0.01 r_z$. A polynomial fit shows a scaling of $\rho \propto \nu^{1.91}$ for the composite system. (c) \ac{fem} simulation of the relative local increase in orbital relaxation times compared to bulk,  $T\tief{1}^0/T\tief{1}^{0,\mathrm{B}}$.}
		\label{fig:afm + simulation}
	\end{figure}
	
	\vspace{10em}
	\textit{Model} -- 
	The groundstate manifold \siv Hamiltonian considered here is given by ($\hbar=1$)
	\begin{align}\label{eq:Hstrain} \nonumber
	\mathrm{H}_\mathrm{SiV} = 
	&\frac{\lambda_\mathrm{SO}}{2} \, \sigma_{y,o} \otimes\sigma_{z,s} \\ 
	&+ \chi_\mathrm{E_{gx}}\epsilon_\mathrm{E_{gx}}\,\sigma_{z,o} \otimes \mathbb{1}_s 
	+\chi_\mathrm{E_{gy}} \epsilon_\mathrm{E_{gy}}\,\sigma_{x,o} \otimes \mathbb{1}_s 
	\end{align}
	where $\sigma_{x,o} = \ket{e_x}\bra{e_y} + \ket{e_x}\bra{e_y}$, $\sigma_{y,o} = -i\ket{e_x}\bra{e_y} +i \ket{e_y}\bra{e_x}$ and $\sigma_{z,o} = \ket{e_x}\bra{e_x} - \ket{e_y}\bra{e_y}$ are orbital and $\sigma_{z,s} = \ket{\uparrow}\bra{\uparrow} - \ket{\downarrow}\bra{\downarrow}$ are spin operators. $\lambda_\mathrm{SO} = \SI{46}{\giga\hertz}$ describes the spin-orbit interaction strength, which lifts the degeneracy by mixing orbital and spin degrees of freedom, to create $\ket{e_{g_\pm}}\ket{\uparrow (\downarrow)} = 1/\sqrt{2}(\ket{e_x} \pm i\ket{e_y})\ket{\uparrow (\downarrow)}$ \cite{heppElectronicStructureSilicon2014}. 
	The strain energies $\chi_r\epsilon_r$ are symmetry adapted linear combinations of strain field components $\epsilon_{ij}$ and strain susceptibilities $\alpha, \beta$ corresponding to irreducible representations $r$ of $D_\mathrm{3d}$ 
	\begin{align}
	\label{eq:salc egx strain}
	\chi_\mathrm{E_{gx}} \epsilon_\mathrm{E_{gx}} &= \alpha\left(\epsilon_{xx}-\epsilon_{yy}\right) + \beta \epsilon_{zx} \\ 
	\label{eq:salc egy strain}
	\chi_\mathrm{E_{gy}} \epsilon_\mathrm{E_{gy}} &= -2\alpha \epsilon_{xy}+\beta\epsilon_{yz} \;,
	\end{align}
	where $\epsilon_{ij}$ are expressed within the defect internal basis \cite{lemondePhononNetworksSiliconVacancy2018}. Diagonal terms involving $\epsilon_\mathrm{A_{1g}}$ are neglected in Eq.\,\eqref{eq:Hstrain}, as they only shift the total energy. As a consequence of the mechanical interaction, the pure spin-orbit states $\ket{e_{g_+}\uparrow}, \ket{e_{g_-}\downarrow}$ and $\ket{e_{g_+}\downarrow}, \ket{e_{g_-}\uparrow}$ with an energy splitting of $\Delta_\mathrm{GS}/2\pi = \lambda_\mathrm{SO}$ will shift in relative energies and also undergo relaxations with rate $\Gamma_1$, see Fig.\,\ref{fig:sketch}. The strain-dependent transition rates from the excited to the ground state are calculated in the above basis by Fermi's golden rule $(T\to0)$
	\begin{align}\nonumber
	\Gamma_1^0 = &\, 2 \pi \sum_n \left(\left| \chi_\mathrm{E_{gx}}\epsilon_{\mathrm{E_{gx}},n}\right|^2 + \left|\chi_\mathrm{E_{gy}}\epsilon_{\mathrm{E_{gy}},n} \right|^2\right) \\ \label{eq:Fermi}
	& \cross \delta \left(\omega_n - \Delta_\mathrm{GS}\right) \; .
	\end{align}
	Here $n$ labels the various coupled system eigenfrequencies. 
	
	The potential local increase in $T^0_1$ compared to bulk $T^{0,\mathrm{B}}_1$ is estimated by utilizing 3D-\ac{fem} to solve for the mechanical eigenfrequencies of a \nd coupled to a diamond substrate.
	The material properties are assumed to be isotropic with mass density $\rho=\SI{3515}{\kilo\gram\per\meter\cubed}$, Young's modulus $E=\SI{1050}{\giga\pascal}$ and Poisson ratio $\nu=0.2$ \cite{lemondePhononNetworksSiliconVacancy2018, lekaviciusDiamondLambWave2019}. Differing material parameters are discussed in the supplementary information.
	
	The \nds within our size range of \SI{30}{}-\SI{100}{nm} exhibit a pronounced cubo-octahedral geometry. The numerically simulated \nd is approximated with a more symmetrical ellipsoidal geometry, whose semi-axis are extracted from an \ac{afm} scan with $(r_x, r_y, r_z)= (40,50,22.5)$nm, see Fig. \,\ref{fig:afm + simulation}(a).
	In addition, slightly immersing the \nd along $z$ into the substrate with a penetration depth $d = \xi r_z$ creates a contact area, where phonons can be exchanged with the bulk reservoir and hence introduce a coupling rate. Fig.\,\ref{fig:afm + simulation}(b) shows the simulated spectral density of states (DOS) $\rho(\nu)$ for the aforementioned \nd coupled to bulk with $\xi = 10^{-2}$, which reveals a quasi localized \nd mode around \SI{75}{\giga\hertz}. Moreover, a polynomial fit (red dashed line in Fig.\,\ref{fig:afm + simulation}(b)) shows that the DOS approximately scales with $\rho(\nu) \propto \nu^{1.91}$ rather than quadratically which is to be expected for a pure bulk-like system in the Debye limit. This indicates contributions from bound surface modes \cite{kepesidisCoolingPhononsPhonons2016, sohnControllingCoherenceDiamond2018}. For details on the DOS evaluation, see supplementary information.
	
	Using the strain susceptibilities $\alpha=\SI{1.3}{\peta\hertz\per Strain}$, $\beta=\SI{1.7}{\peta\hertz\per Strain}$ in Eq.\,(\ref{eq:salc egx strain}) \& (\ref{eq:salc egy strain}), as well as the simulated strain fields $\epsilon_{ij,n}(\vec{r})$ of the coupled system, where the \siv high-symmetry axis is assumed along $z$, Eq.\,(\ref{eq:Fermi}) is evaluated at each position in the $xz$-plane \cite{meesalaStrainEngineeringSiliconvacancy2018}. For this the delta-distributions $\delta_n = \delta(\omega_n - \Delta_\mathrm{GS})$ are approximated with Lorentzians and the correspondingly simulated eigenfrequencies $\omega_n$ as well as decay rates $\gamma_n = \omega_n / Q_n$ are used, where $Q_n$ is the respective mechanical quality factor. 
	Fig.\,\ref{fig:afm + simulation}(c) shows the orbital lifetime ratio $T^0_1 / T_1^{0,\mathrm{B}}$, where $T_1^{0,\mathrm{B}} \approx \SI{209}{\nano\second}$ is obtained by averaging $T^0_1$ over the bulk which is in good agreement with the analytical $T_1^{0,\mathrm{B}} \approx \SI{233}{\nano\second}$. Note that Fig.\,\ref{fig:afm + simulation}(c) only shows a close-up of the \nd in the $xz$-plane. Further details on the whole simulation are presented in the supplementary information. 
	\begin{figure}[t]
		\centering
		\begin{tikzpicture}
		\draw (0, 0) node[inner sep=0] {\includegraphics[scale=1]{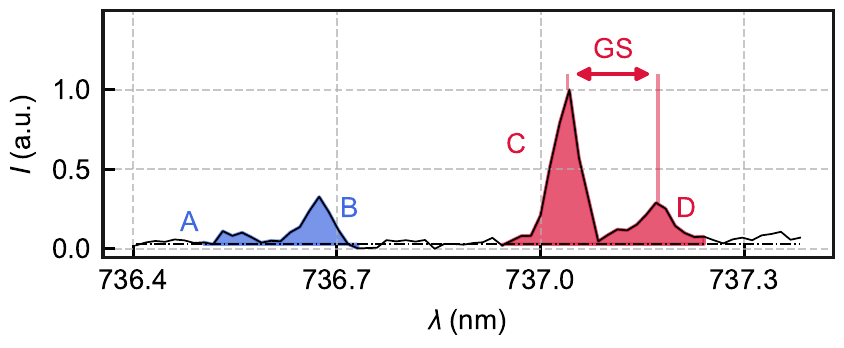}};
		\draw (-115pt, 50pt) node {(a)};
		\end{tikzpicture}
		\begin{tikzpicture}
		\draw (0, 0) node[inner sep=0] {\includegraphics[scale=1]{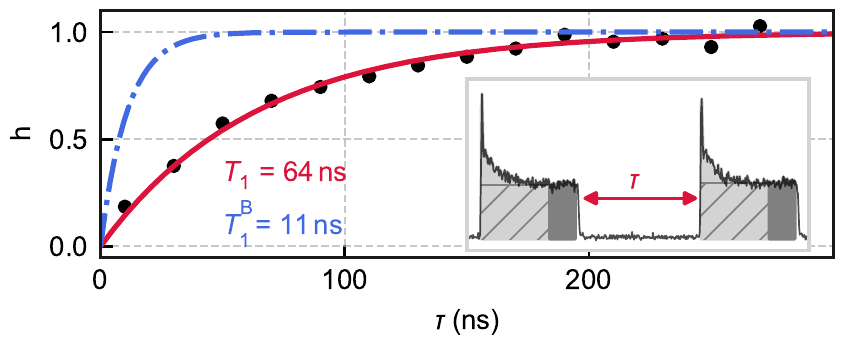}};
		\draw (-115pt, 50pt) node {(b)};
		\end{tikzpicture}
		\caption{(a) Photo-luminescence intensity $I$ as a function wavelength $\lambda$ revealing the four-line structure of a \siv in an \nd with a \gss of \SI{73}{\giga\hertz}. The temperature of the \siv is determined using Boltzmann distributed populations and summing over transition A+B and C+D. (b) Peak heights $h$, determined from the fluorescence of a resonant pulse sequence, as a function of inter-pulse delay $\tau$. The solid red line shows a numerical fit revealing $T_1 = \SI{64}{\ns}$, whereas the dash-dotted blue line depicts a correspondingly theoretical bulk recovery with $T_1^\mathrm{B} = \SI{11}{\ns}$ as a comparison. The inset illustrates the extraction of the peak heights with the extrapolation of the stationary count rate (horizontal solid line) by the last \SI{50}{ns} countrace (dark grey area).
		}
		\label{fig:spectrum_and_t1}
	\end{figure}
	\begin{figure}[t]
		\centering
		\includegraphics[scale=1]{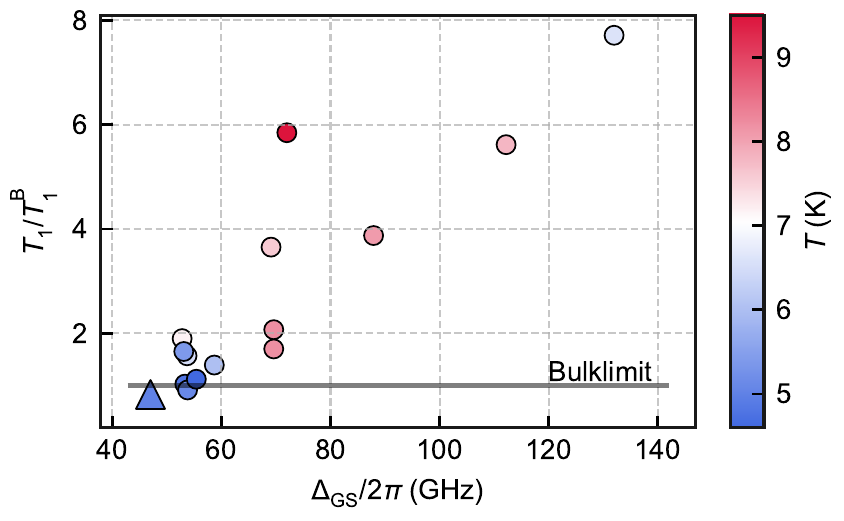}
		\caption{Orbital lifetime extension through comparison of measured $T_1$ and analytical bulk-limit $T_1^{\mathrm{B}}$ for different \gss $\Delta_\mathrm{GS}$ and temperatures $T$. The triangularly shaped marker represents a measured $T_1 = \SI{38}{\nano\second}$ in bulk from reference \cite{jahnkeElectronPhononProcesses2015}.}
		\label{fig:ratio}
	\end{figure}
	The already discussed presence of surface acoustic waves, bound to the diamond interface, also increases strain and hence leads to a reduction in $T_1^0$ \cite{maityCoherentAcousticControl2020}. At the contact area, where strain fields are largest, the relaxation rate is maximal with two orders of magnitude smaller decay time than the average $T_1^{0,\mathrm{B}}$. However, the current configuration reaches a maximum of $T^0_{1,\mathrm{max}} / T_1^{0,\mathrm{B}} \approx 120$ at the center of the \nd. Even in the upper half of the \nd, relaxation time is enhanced by approximately an order of magnitude. As a rough estimate, $T_{1,\mathrm{max}}^{0}$ would extend the orbital lifetime at cryogenic temperatures of $T=\SI{5}{\kelvin}$ to $T_{1,\mathrm{max}} = \coth(\hbar \Delta_\mathrm{GS}/ 2 k_\mathrm{B} T) T_{1,\mathrm{max}}^{0} \approx \SI{5}{\micro\second}$.

	\textit{Measurements} -- \Timeone is measured for 14 \siv in \nds at temperatures ranging from \SI{4.6}{K} up to \SI{9.5}{K}. The \nds are coated onto a diamond substrate, selected for high thermal conductivity. Using an \ac{afm}, the average size of the \nds is determined to be \SI{30}{\nano\meter}. 
	The temperature of each \siv is evaluated by dividing the sum of normalized counts in transition C and D ($C_\mathrm{C+D}$) by the respective counts in A and B ($C_\mathrm{A+B}$), which reflects the relative population in the lower and upper excited state, see Fig.\,\ref{fig:spectrum_and_t1}(a). Assuming a Boltzmann distribution for the latter, the temperature can be calculated with
	\begin{equation}
	\label{eq-temp}
	\frac{C_\mathrm{C+D}}{C_\mathrm{A+B}}= \exp\left(-\frac{\hbar \Delta\tief{ES}}{k_\mathrm{B} T}\right) .
	\end{equation} 
	The \ess is determined from the spectrum of an off-resonantly excited \siv. Fig.\,\ref{fig:spectrum_and_t1}(a) shows an example \nd spectrum.
	
	Resonant excitation of transition C is achieved with a tunable single-frequency Ti:sapphire laser and fluorescence from the phonon-sideband is detected using a \SI{750}{\nano\meter} longpass filter in front a single-photon counter to block the laser. Additionally, a weak \SI{532}{\nano\meter} laser is utilized to stabilize the emission. 
	\Timeone is measured with a tailored pulse-sequence, consisting of several \SI{200}{\nano\second}-long pulses with an increasing inter-pulse delay $\tau$ \cite{hausslerPreparingSingleSiV2019}. 
	The peak heights are extracted by summing up all counts within each pulse and subtracting the stationary ones. The latter is estimated by extrapolating a stationary count rate using the last \SI{50}{ns} and multiplying it by the corresponding pulse length, as depicted in the inset of Fig.\,\ref{fig:spectrum_and_t1}(b).
	The peak heights $h$ are then fitted with
	\begin{equation}
	h(\tau)\propto 1-\exp\left(-\frac{\tau}{T\tief{1}}\right) \;,
	\label{eq-t1}
	\end{equation}
	revealing a \Timeone of \SI[separate-uncertainty = true]{64(3)}{\ns} for the \siv measured in Fig.\,\ref{fig:spectrum_and_t1}(b). 
	
	To compare our measured values for \Timeone at different temperatures $T$ and \gss $\Delta_\mathrm{GS}$ to the corresponding bulk time $T_1^\mathrm{B}$, an analytical expression 
	\begin{align}
	\label{eq:T1bulk}\nonumber
	\left(T_1^\mathrm{B}\right)^{-1} = &\frac{h(\alpha^2+(\beta/2)^2)}{\pi \rho} \left( \frac{1}{5v_t^5} + \frac{2}{15v_l^5} \right)\left(\frac{\Delta_\mathrm{GS}}{2\pi}\right)^3\\
	&\times \coth\left(\frac{\hbar\Delta_\mathrm{GS}}{2k\tief{B}T}\right) \, ,
	\end{align}
	with
	\begin{align}
	\nonumber
	v_t = \sqrt{\frac{E}{\rho} \, \frac{1}{2(1+\nu)}}\, \quad
	v_l = \sqrt{\frac{E}{\rho} \, \frac{1-\nu}{(1+\nu)(1-2\nu)}}
	\end{align}
	is used. Eq.\,(\ref{eq:T1bulk}) can be derived from Eq.\,(\ref{eq:Fermi}) using a long-wavelength approximation \cite{albrechtCouplingNitrogenVacancy2013, kepesidisCoolingPhononsPhonons2016}. 
	For the \Timeone measurement shown in Fig.\,\ref{fig:spectrum_and_t1}(b), the corresponding \nd temperature of $T=\SI{9.5}{\kelvin}$ and \gss $\Delta_\mathrm{GS}/2\pi = \SI{72}{\giga\hertz}$ together with Eq.\,\eqref{eq:T1bulk} are used to calculate a bulk orbital relaxation time of $T_1^\mathrm{B} = \SI{11}{ns}$. Hence, an increase by a factor of $\SI{64}{\ns}/\SI{11}{\ns}\approx 6$ for this \nd is achieved.
	The procedure is repeated for all other 13 \nds and the ratio $T_1 / T_1^\mathrm{B}$ is depicted in Fig.\,\ref{fig:ratio} as a function of \gss. The color of each dot shows the temperature of the respective \siv. For a \gss of \SI{132}{\giga\hertz} a prolongation of the orbital lifetime of up to 8 can be observed. The triangularly shaped marker represents a measured \Timeone in bulk which was conducted under similar experimental conditions \cite{jahnkeElectronPhononProcesses2015}. 
	The above calculated $T_1^\mathrm{B}$ is a conservative estimation of the upper limit reachable in bulk-diamond in the presence of resonant, single-phonon dominated orbital relaxation. To our knowledge experimentally measured $T_1^\mathrm{B}$ in the low-strain regime do not reach the theoretical limit \cite{rogersAllOpticalInitializationReadout2014, beckerAllOpticalControlSiliconVacancy2018}.
	
	\textit{Discussion} -- The results in Fig.\,\ref{fig:ratio} suggest an inverse correlation between $T_1/T_1^\mathrm{B}$ and \nd size. On the one hand, reducing the size shifts the \nds mechanical eigenfrequenices to higher energies \cite{albrechtCouplingNitrogenVacancy2013}, which also enhances the respective modes strain field \cite{lemondePhononNetworksSiliconVacancy2018,jahnkeElectronPhononProcesses2015}. As a result a \siv can couple to these fields, which increases its \gss. Consequential, thermally excited phonons from the surrounding \nd or bulk are exponentially less likely populated, thus reducing \Gammaone. On the other hand, smaller \siv host geometries exhibit reduced contact areas between the \nd and the substrate which then suppress coupling to the bulk phonon bath, hence also reducing \Gammaone. The supplementary information provides further simulations with a different geometry compared to Fig.\,\ref{fig:afm + simulation}, various contact areas and ground-state splittings. 
	Geometrical decoupling becomes apparent, when looking at \nds with $\Delta_\mathrm{GS}/2\pi \approx \SI{70}{\giga\hertz}$ in Fig.\,\ref{fig:ratio}. The similar GS splittings indicate correspondingly comparable \nd geometries. However, the one with a maximal temperature of $T=\SI{9.5}{\kelvin}$, exhibits an orbital lifetime ratio of $T_1/T_1^\mathrm{B} \approx 6$, which is largest for those \nds under consideration. This might reveal a preferential orientation of the \nd, whose contact area with the substrate is smallest and thus also isolated best. As a result the phonon exchange with the substrate is reduced which reduces \Gammaone and increases temperature, too. 
	The discrepancy between simulated orbital lifetime ratios and the ones which are determined from measurements is attributed to an unfavorable interplay of different mechanisms, leading to non-trivial coupling rates to the bulk. These broaden the \nd resonances and thus bring their local phononic spectral densities closer to the one of bulk which limit the maximally achievable $T_1/T_1^\mathrm{B}$. Firstly, although the \nds sizes are measured to be smaller than the simulated one, a contact area largely exceeding the one used in the simulation strongly enhances phonon exchange with the bulk. Secondly, the mechanical impedances of the \nd and the diamond base substrate are similar, allowing for a high phonon transmission, which also increases coupling to the substrate.
	
	\textit{Outlook} -- The experimental results presented in this work show that the \siv orbital lifetime is extended by a factor of 8 when incorporated into a small, tailored diamond host as compared to a \siv in bulk-diamond. In order to suppress the orbital relaxation further, nano-manipulation techniques utilizing an \ac{afm} can be used to rotate the \nds, thereby lowering the area of contact which leads to a more optimal isolation \cite{hausslerPreparingSingleSiV2019, rogersSingleSiCenters2019}. In addition, using a substrate material which is engineered to suppress phonons in the relevant spectral range also reduces the impedance matching between the former and the \nd, which yields a better isolated host. For this purpose, a phononic crystal with a suitable bandgap could be used  \cite{lekaviciusDiamondLambWave2019}. Ultimately, the concepts could be applied to experiments with levitated \nds \cite{hoangElectronSpinControl2016, juanCooperativelyEnhancedDipole2017, frangeskouPureNanodiamondsLevitated2018}.
	
	The \siv spin dephasing time \Timetwostar is mainly limited by orbital relaxation processes involving single phonons at temperatures around \SI{5}{\kelvin} \cite{rogersAllOpticalInitializationReadout2014}, with $T_2^\star \approx 2 T_1$ \cite{pingaultCoherentControlSiliconvacancy2017, beckerAllOpticalControlSiliconVacancy2018}. Hence, extending $T_1$ to the predicted \SI{5}{\micro\second} opens up the possibility to coherently control the \siv spin without relying on a dilution refrigerator, since mechanical decoupling from thermal phonons was shown for temperatures below \SI{500}{\milli\kelvin} \cite{sukachevSiliconVacancySpinQubit2017}. Thus in this regime, \Timetwostar is no longer limited by single phonon processes.
	
	Moreover, due to the \nds integration capability into photonic structures, passive photonic and phononic properties could be merged and leveraged to achieve coherent spin-photon control \cite{fehlerPurcellenhancedEmissionIndividual2020}.
	
	\section{Acknowledgments}
	The project was funded by the Deutsche Forschungsgemeinschaft (DFG, German Research Foun dation) in project 398628099, the Baden-Württemberg Stiftung in project Internationale Spitzenforschung and IQst. MWD acknowledges support from the Australian Research Council (DE170100169). Experiments performed for this work were operated using the Qudi software suite \cite{binderQudiModularPython2017}. 

%

\end{document}